\documentclass{PoS}

\title{High-precision measurements of extensive air showers with the SKA}

\ShortTitle{High-precision measurements of extensive air showers with the SKA}

\newcommand{\manchester}{School of Physics \& Astronomy, Univ.\ of Manchester, M13 9PL, United Kingdom}
\newcommand{\nijmegen}{Dept.\ of Astrophysics/IMAPP, Radboud Univ.\ Nijmegen, 6500 GL Nijmegen, The Netherlands}
\newcommand{\astron}{Netherlands Institute for Radio Astronomy (ASTRON), 7990 AA Dwingeloo, The Netherlands}
\newcommand{\karlsruhe}{IKP, Karlsruher Institut f\"ur Technologie, Postfach 3640, 76021 Karlsruhe, Germany}
\newcommand{\erlangen}{ECAP, Univ.\ of Erlangen-Nuremberg, 91058 Erlangen, Germany}
\newcommand{\groningen}{Kernfysisch Versneller Instituut, Univ.\ of Groningen, 9747 AA Groningen, The Netherlands}
\newcommand{\subatech}{Subatech, 4 rue Alfred Kastler, 44307 Nantes cedex 3, France}
\newcommand{\nancay}{Station de radioastronomie de Nan\c cay, Observatoire de Paris, CNRS/INSU, Nan\c cay, France}
\newcommand{\atnf}{CSIRO Astronomy \& Space Science, NSW 2122, Australia}
\newcommand{\karlsruheekp}{EKP, Karlsruher Institut f\"ur Technologie, Kaiserstr. 12, 76131 Karlsruhe, Germany}
\newcommand{\vub}{Astrophysical Institute, Vrije Universiteit Brussel, Pleinlaan 2, 1050 Brussels, Belgium}
\newcommand{\iihe}{Interuniversity Institute for High-Energy, Vrije Universiteit Brussel, Pleinlaan 2, 1050 Brussels, Belgium}
\author{
 \speaker{T.~Huege}$^1$, J.D.~Bray$^2$, S.~Buitink$^3$, 
 R.~Dallier$^{4,5}$, R.D.~Ekers$^6$, H.~Falcke$^{7,8}$, A.~Haungs$^1$, C.W.~James$^9$, L.~Martin$^{4,5}$, B.~Revenu$^4$, O.~Scholten$^{10,11}$,  F.G.~Schr\"oder$^1$ and A.~Zilles$^{12}$\\
 $^1$\karlsruhe \\
 $^2$\manchester \\
 $^3$\vub \\
 $^4$\subatech \\
 $^5$\nancay \\
 $^6$\atnf \\
 $^7$\nijmegen \\
 $^8$\astron \\
 $^9$\erlangen \\
 $^{10}$\groningen \\
 $^{11}$\iihe \\
 $^{12}$\karlsruheekp \\
 E-mail: \email{tim.huege@kit.edu}
}
               
\abstract{As of 2023, the Square Kilometre Array will constitute the 
world's largest radio telescope, offering unprecedented capabilities 
for a diverse science programme in radio astronomy. At the same time, 
the SKA will be ideally suited to detect extensive air showers 
initiated by cosmic rays in the Earth's atmosphere via their  
radio emission. With its very dense and uniform antenna spacing in a 
fiducial area of one km$^2$ and its large bandwidth of 50-350 
MHz, the low-frequency part of the SKA will provide very precise 
measurements of individual cosmic ray air showers. These precision measurements
will allow detailed studies of the mass composition of cosmic rays in the
energy region of transition from a Galactic to an extragalactic origin.
Also, the SKA will facilitate three-dimensional ``tomography'' of the
electromagnetic cascades of air showers, allowing the study of particle
interactions at energies beyond the reach of the LHC. Finally, studies of possible connections between air showers
and lightning initiation can be taken to a new level with the SKA.
We discuss the science potential of air shower detection with the SKA 
and report on the technical requirements and project status.}

\FullConference{The 34th International Cosmic Ray Conference,\\
		30 July- 6 August, 2015\\
		The Hague, The Netherlands}

\begin{document}

\section{Introduction}

Radio detection of extensive air showers has made great progress in 
the past decade \cite{huegereview}. Currently, the second generation 
of digital radio detection experiments is in full swing.
Two complementary strategies are being followed. One concept is to instrument 
large areas, spacing antennas as sparsely as possible, as is done in the Auger 
Engineering Radio Array (AERA) \cite{schulzaera} with 150 antennas 
spanning an area of $\approx 17$~km$^{2}$. The other approach is to observe air showers with a very dense array of 
antennas on a fairly small area, as is the case in LOFAR \cite{lofar}, 
with several hundred antennas on an area of $\approx 0.2$~km$^{2}$.

With the latter approach, very detailed information on individual air 
showers can be gathered. Matching state-of-the-art simulations of the 
radio signal from extensive air showers \cite{coreas} to measured LOFAR data 
allows in particular an accurate reconstruction of the depth of shower 
maximum ($X_{\mathrm{max}}$) of extensive air showers. The average 
$X_{\mathrm{max}}$ uncertainty of 
LOFAR measurements has been quantified at 17~g/cm$^{2}$ \cite{lofarxmax}.

As of 2020, the low-frequency part of the first phase of the Square Kilometre Array 
(SKA-low) will go into operation in Australia, its completion being
envisaged for 2023. Already in this first 
phase, the core of SKA-low will be comprised of approximately
60,000 dual-polarized antennas deployed in a 
circle with 750~m diameter. With moderate engineering changes to 
the SKA-low baseline design, these antennas can be used for 
high-precision measurements of extensive air showers, in a fiducial 
area of roughly one km$^{2}$. The SKA focus group on high-energy cosmic 
particles\footnote{http://astronomers.skatelescope.org/home/focus-groups/high-energy-cosmic-particles/} has undertaken the effort to work with the SKA organization towards enabling cosmic ray detection with the SKA using extensive air showers (this article) as well as lunar detection \cite{ska-lunar}.

In this article, we shortly discuss the science that could be carried out with 
SKA-low\footnote{Throughout this article, we use ``SKA-low'' to refer to the 
Phase 1 instrument only. Phase 2 is expected to be completed by 2030 and anticipated to 
be at least four times the size of Phase 1.} once enabled for air-shower detection. Then, we discuss the 
needed engineering changes to enable SKA-low as a cosmic-ray detector. 
Finally, we provide a glimpse at the potential of air 
shower detection with SKA-low on the basis of simulation studies.

\section{Science potential}

The key point of the radio detection of extensive air showers with 
the SKA will be precision. The high number of antennas measuring 
individual air showers, their homogeneous distribution in the SKA-low 
core and the large instantaneous bandwidth from 50 to 350~MHz will allow measurements 
with unprecedented quality. Measurements of $\sim 10,000$ showers above an 
energy of 10$^{17}$~eV are expected per year. With these high-fidelity 
data, the following scientific questions can be addressed:

\subsection{Transition from Galactic to extragalactic cosmic rays}

\begin{figure}
\centering
\includegraphics[width=0.45\textwidth]{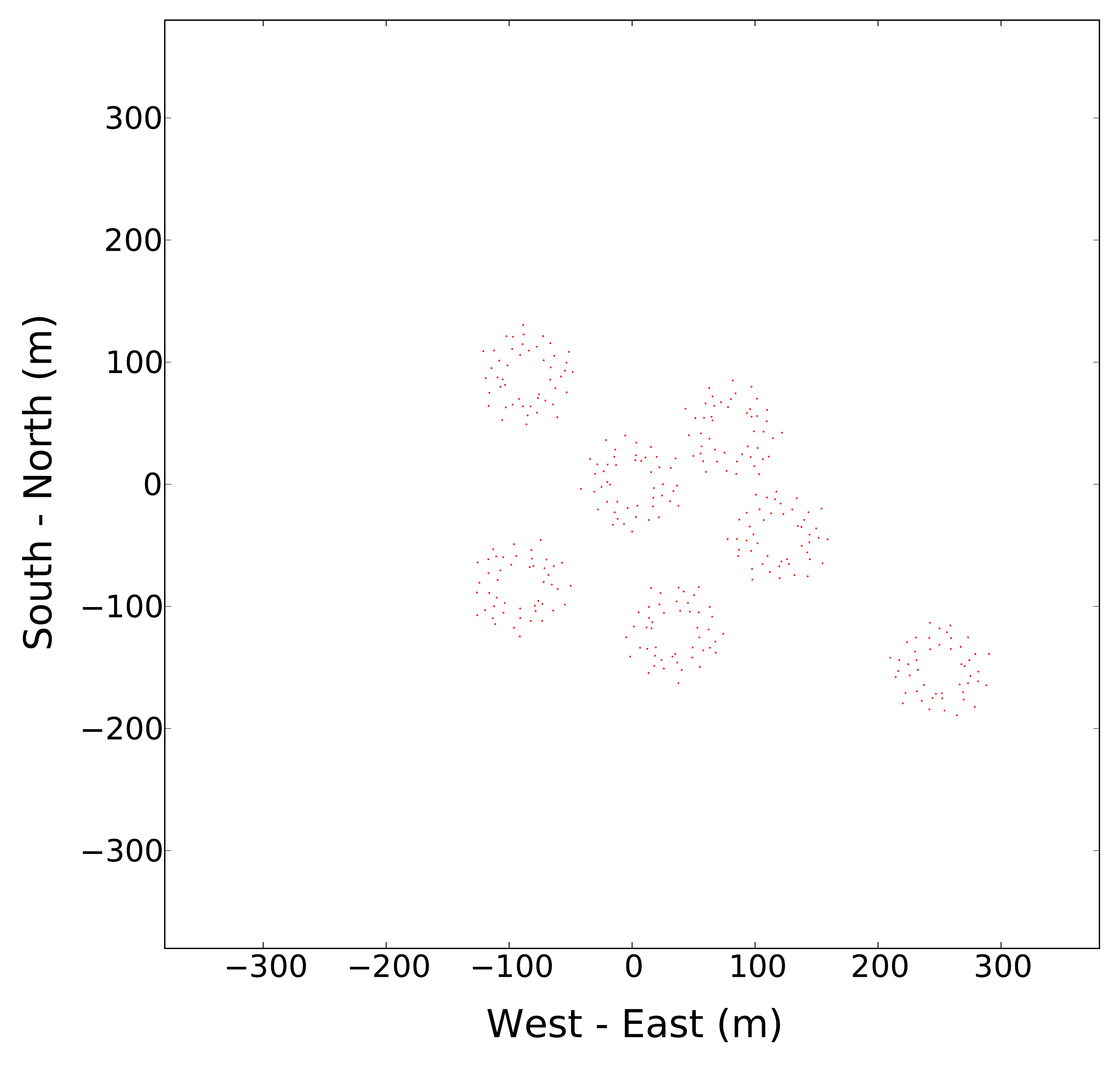}
\includegraphics[width=0.45\textwidth]{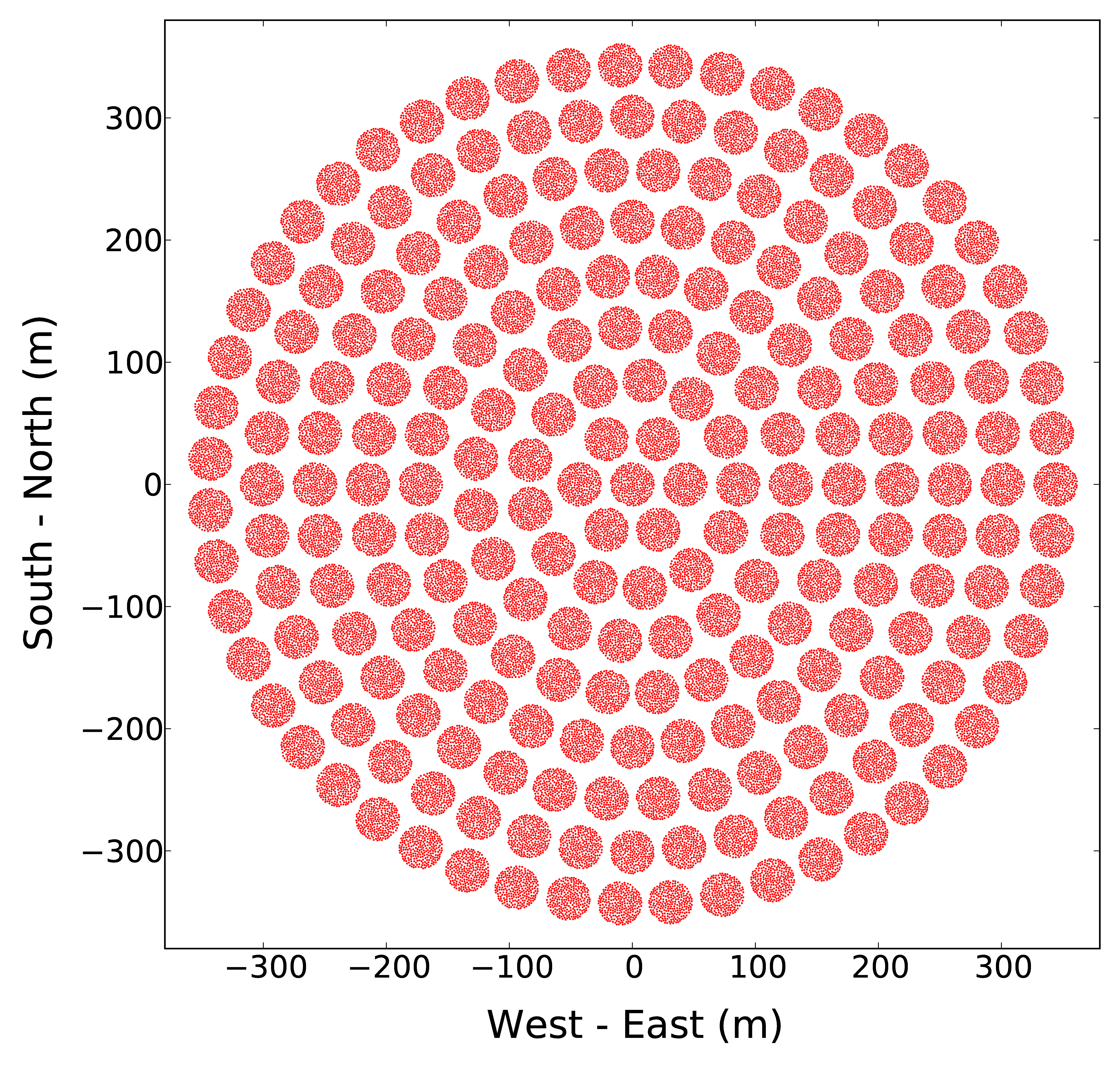}
\caption{Illustrations of the antenna distributions typically used for 
air shower observations in the LOFAR core (left) and the SKA-low core (right). Each point represents one 
dual-polarized antenna. The SKA-low is organized in stations with a 
diameter of 35~m comprised of 256 antennas each.\label{fig:detector_layouts}}
\end{figure}

With SKA-low measurements the mass composition in the energy range from $\gtrsim 10^{16}$~eV 
up to $\gtrsim 10^{18}$~eV can be probed with unprecedented 
resolution. This could in particular shed light on the transition from 
Galactic to extragalactic cosmic rays. With LOFAR, the depth of 
shower maximum has been successfully reconstructed with a mean accuracy of 
17~g/cm$^{2}$. This resolution is limited by the incomplete sampling 
of the radio emission footprint with the inhomogeneous antenna distribution 
of LOFAR, see Fig.\ \ref{fig:detector_layouts} (left). 
%
%
In the case of individual air shower events for which the important features 
in the radio emission footprint (lateral slope, asymmetry, Cherenkov 
bump) are well-sampled in spite of the inhomogeneous distribution of 
antennas LOFAR can achieve a precision on the depth of shower maximum as low as 8~g/cm$^{2}$.
In comparison, SKA-low will provide much more detailed and much more
homogeneous measurements of the radio emission footprint, as is 
illustrated in Fig.\ \ref{fig:detector_layouts} (right). With this and 
other improvements, we expect to achieve a \emph{mean} $X_{\mathrm{max}}$ resolution 
below 10~g/cm$^{2}$, significantly better than the $X_{\mathrm{max}}$ resolution of any 
other technique available to date. Also, these high-precision 
measurements will be carried out with a 
near-100\% duty cycle. Identification of individual elements in the 
cosmic ray flux, in particular the separation between proton showers 
and other nuclei, will thus become feasible.

\subsection{Air shower physics beyond the LHC scale}

Uncertainties in particle interactions at energies beyond those 
reached by the LHC and in the regime of extreme forward kinematics still limit the interpretation of air shower 
data in many ways. Measurements of the moments of the 
$X_{\mathrm{max}}$ distributions can be used to probe 
high-energy particle interactions, in particular with respect to the 
proton-air cross section, secondary particle multiplicity, elasticity 
and pion charge ratio \cite{ulrich}. With its very accurate 
$X_{\mathrm{max}}$ determination, a much cleaner separation between 
proton and helium nuclei than previously achieved should be feasible. 
SKA-low in particular has the potential to perform measurements of 
the proton-air cross section \cite{AugerCC} with small systematic uncertainties.

Radio signals from extensive air showers, however, contain much more information 
than just the depth of shower maximum. The entire longitudinal 
evolution of the air shower is encoded in the radio signal. Using near-field 
interferometric analysis techniques, it should be possible to perform 
a three-dimensional ``tomography'' of the electromagnetic cascade of 
extensive air showers, potentially allowing very detailed studies of air-shower 
physics. The large number of antennas, their homogeneous spacing and 
the large instantaneous bandwidth reaching up to 350~MHz will be key to realizing this 
approach.

\subsection{Cosmic rays and thunderstorms}

Already with LOPES it has been demonstrated that atmospheric electric 
fields during thunderstorms influence radio emission from extensive 
air showers \cite{lopesthunder}. With LOFAR measurements, it was 
recently demonstrated that the radio 
detection of extensive air showers during thunderstorm conditions can 
probe the atmospheric electric fields in thunderclouds 
\cite{lofarthunder}. Such measurements provide the unique possibility 
of studying the conditions leading to lightning 
initiation in situ. SKA-low will take such measurements to the next 
level, allowing very detailed studies of thundercloud 
systems. It has also been hypothesized that the ionisation introduced 
by extensive air showers could be at the heart of lightning 
initiation, e.g., via the so-called ``relativistic runaway breakdown'' 
mechanism \cite{RunawayBreakdown}, or by providing seed electrons that initiate discharges in the enhanced fields near hydrometeors \cite{LightningInception}. With simultaneous measurements of radio pulses emitted by lightning strikes and the radio emission from 
coincident extensive air showers, SKA-low could also shed new light on this 
question.

\subsection{Precision studies of air shower radio emission}

Finally, the high-fidelity data of SKA-low can be used to study the 
radio emission from extensive air showers itself with unprecedented 
precision. This way it will be possible to validate state-of-the-art 
simulation codes \cite{coreas,zhaires} at an extreme level of detail, and at much higher 
frequencies than possible in experiments existing so far. As these 
simulation codes are used widely as the basis for design studies and event 
reconstruction strategies, the scientific community would profit enormously from 
a high-quality validation of the calculations.

\section{Engineering changes}

SKA-low was originally foreseen as an observatory 
with its baseline design optimized to various radio-astronomical 
observation modes. The baseline design is thus not suited to perform measurements of 
extensive air showers. Here, we discuss the engineering changes that 
are needed to enable SKA-low as a detector for extensive air showers.

\subsection{Buffering of radio data}

SKA-low is an aperture-synthesis array with individual 
radio antennas sensitive to the full sky. The signals measured by 
individual antennas are transmitted to a central location via analog, optical 
``RF over fibre'' links. There, they are digitized and processed 
further to form beams in certain directions on the sky as governed by 
the ongoing astronomical observations.

As extensive air showers arrive randomly on the sky, without a priori 
knowledge of direction and arrival time, air shower detection needs access to the 
non-beamformed data of the individual antennas. The individual antenna 
signals thus need to be buffered for long enough until a trigger 
marking the arrival of an extensive air shower has been received and 
initiates the buffer readout. (We 
foresee a particle detector array to deliver such a trigger within 
10~ms, see next section.) The most important engineering change to 
SKA-low is thus the installation of suitable buffering capacity at 
individual antennas. These need to store the raw, unprocessed signal 
for each polarization of each antenna, with a dynamic range of at 
least 8 bits, ideally 12 bits. The sampling rate of SKA-low is 
800~MSPS. Buffering for air shower detection should be completely 
independent of other buffering schemes as, e.g., foreseen for the detection of 
astronomical transients, and should be working continuously, 
independent of other ongoing observations. In this way, radio detection 
will be available with a near 100\% duty-cycle, and will be completely 
commensal to astronomical observations with the SKA.

\subsection{Triggering with particle detectors}

The most reliable way to provide an efficient and pure trigger for the 
readout of the antenna buffers is a particle detector array. 
Radio-only triggering would require very powerful real-time analysis 
of the SKA-low data, and would yield a higher detection 
threshold and probably a high rate of false positive triggers.

The particle detector should become efficient at $\gtrsim 10^{16}$~eV 
and deliver a trigger within 10~ms. Its main 
purpose is to provide a reliable trigger for the readout of the radio 
antennas. The reconstruction of the cosmic ray parameters, in particular 
geometry, energy, and depth of shower maximum, will rely mostly on the 
high-fidelity radio measurements. A certain separation in the measurement of the 
electromagnetic and muonic component of the particle cascade will be 
possible by exploiting the timing and the spatial distribution of the 
signals measured with the particle detectors. Possibly, a subset of the particle detectors could also be 
buried to shield the electromagnetic component of the cascade.

\begin{figure}
\centering
\includegraphics[width=0.6\textwidth]{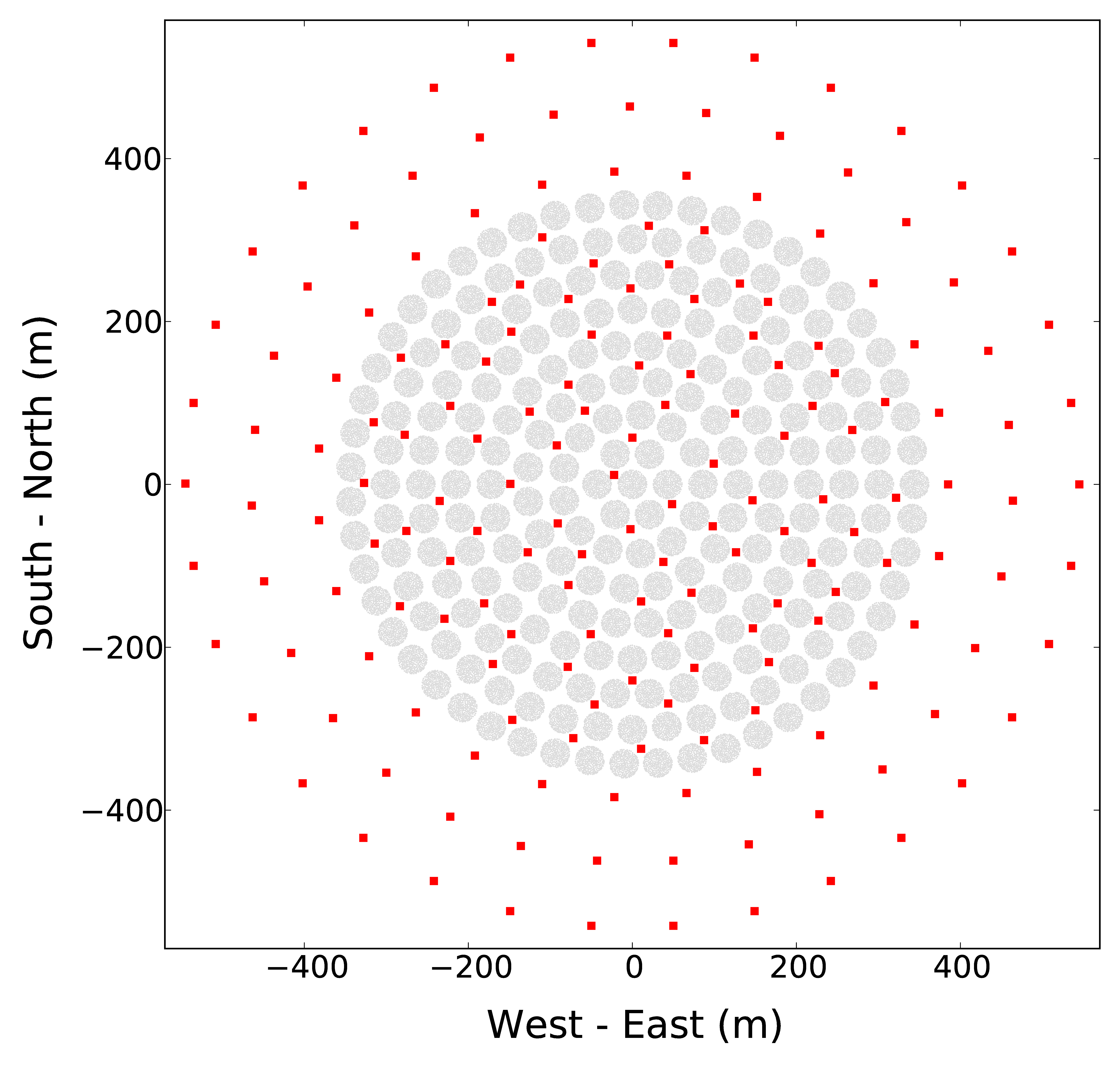}
\caption{Possible layout of the particle detector array. Red squares: 
180 particle detector units, gray circles: 230 SKA-low antenna 
stations with 256 individual antennas each. The exact layout is still under review.}
\label{fig:pd_grid}
\end{figure}

One option for the installation of such a particle detector array is 
to re-use scintillation detectors from the now-dismantled KASCADE 
array \cite{kascade}. A total of 180 high-quality scintillator stations
consisting of 4 tiles each with dimensions of 90~x~90~x~3~cm$^{3}$ could be 
used to instrument a circular area with a diameter of 1.1~km, as is 
shown in Fig.\ \ref{fig:pd_grid}. Air 
showers impacting within this fiducial area can be reconstructed, even if their 
core is outside the SKA-low antenna core.

Particular care has to be taken not to introduce any radio-frequency 
interference in SKA-low via the particle detectors, as they will be deployed in 
very close proximity to the radio antennas. Thorough shielding of the 
detectors, and the use of silicon photomultipliers instead of 
photomultiplier tubes operated at high voltage, will ensure that 
no impulsive radio signals emanate from the detectors. Furthermore, 
data readout will be performed with the same ``RF over fibre'' optical 
links as used for antennas, i.e., digitization and coincidence searches
will take place in a central location, and no clock needs to be distributed
to the particle detectors. The only electrical connection will be for 
power.

The particle detector signals will have to be monitored continuously for 
coincidences and a sophisticated trigger logic will have to be 
implemented. The goal is to achieve detection unbiased by primary 
particle type at energies $\gtrsim 10^{16}$~eV, yet keep the event 
rate to at most one per minute so as to not put too big a strain on the 
bandwidth needed to read out the buffers. On a trigger being received, the buffers 
will be frozen to read out a window of 50~$\mu$s or less, and the data 
analysed offline.

\section{Initial simulation studies}

\begin{figure}
\centering
\includegraphics[width=0.48\textwidth]{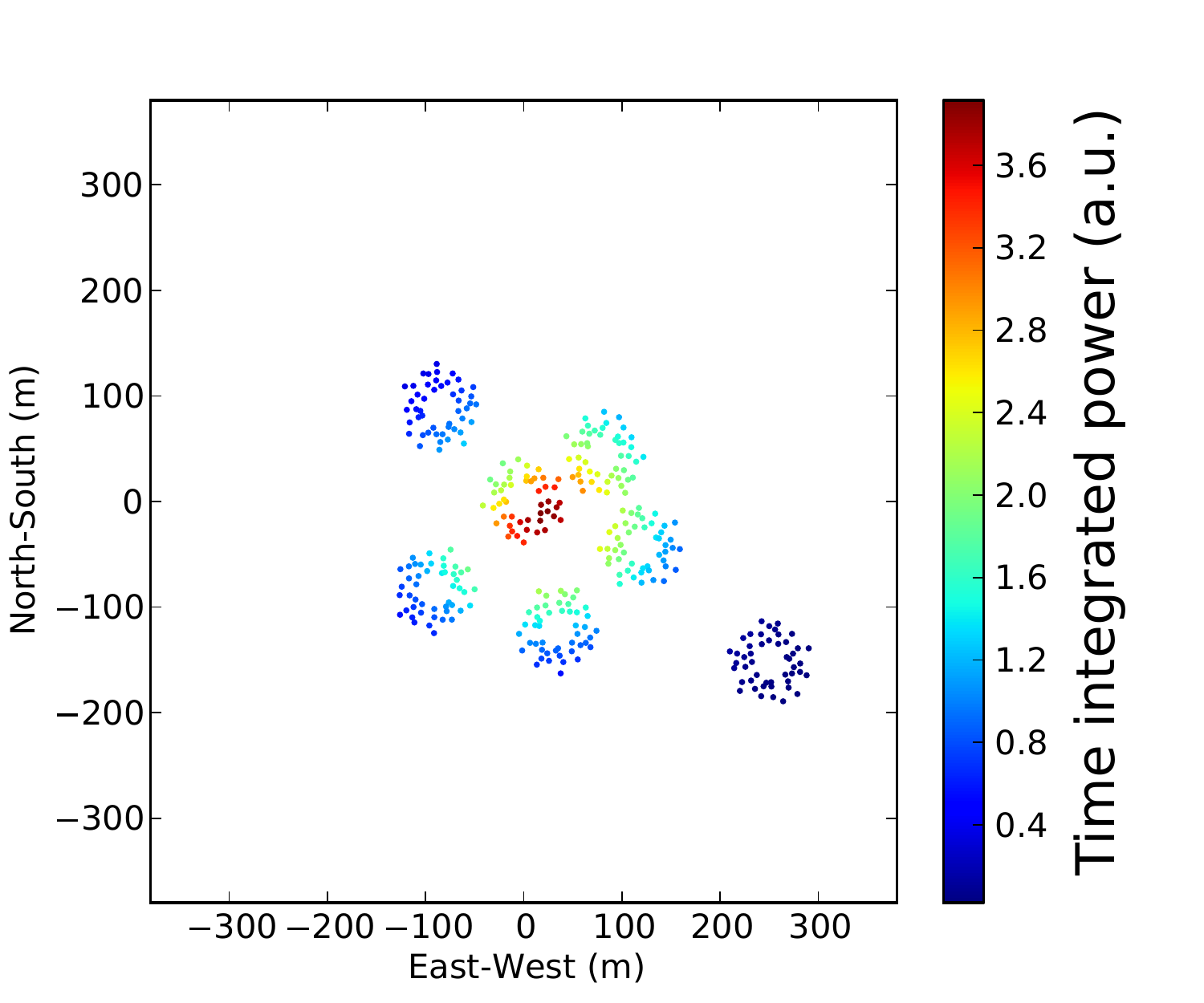}
\includegraphics[width=0.48\textwidth]{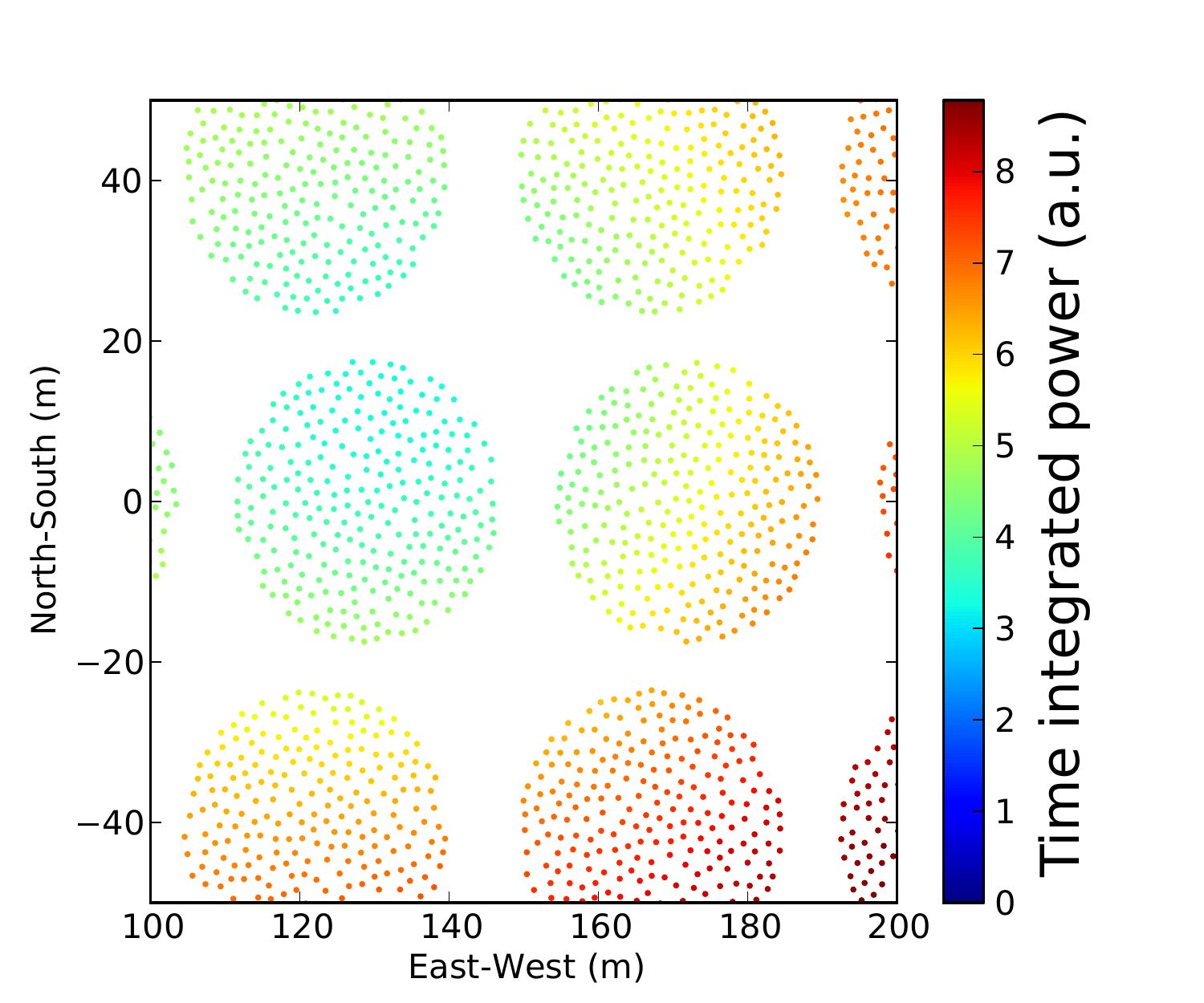}
\includegraphics[width=0.99\textwidth]{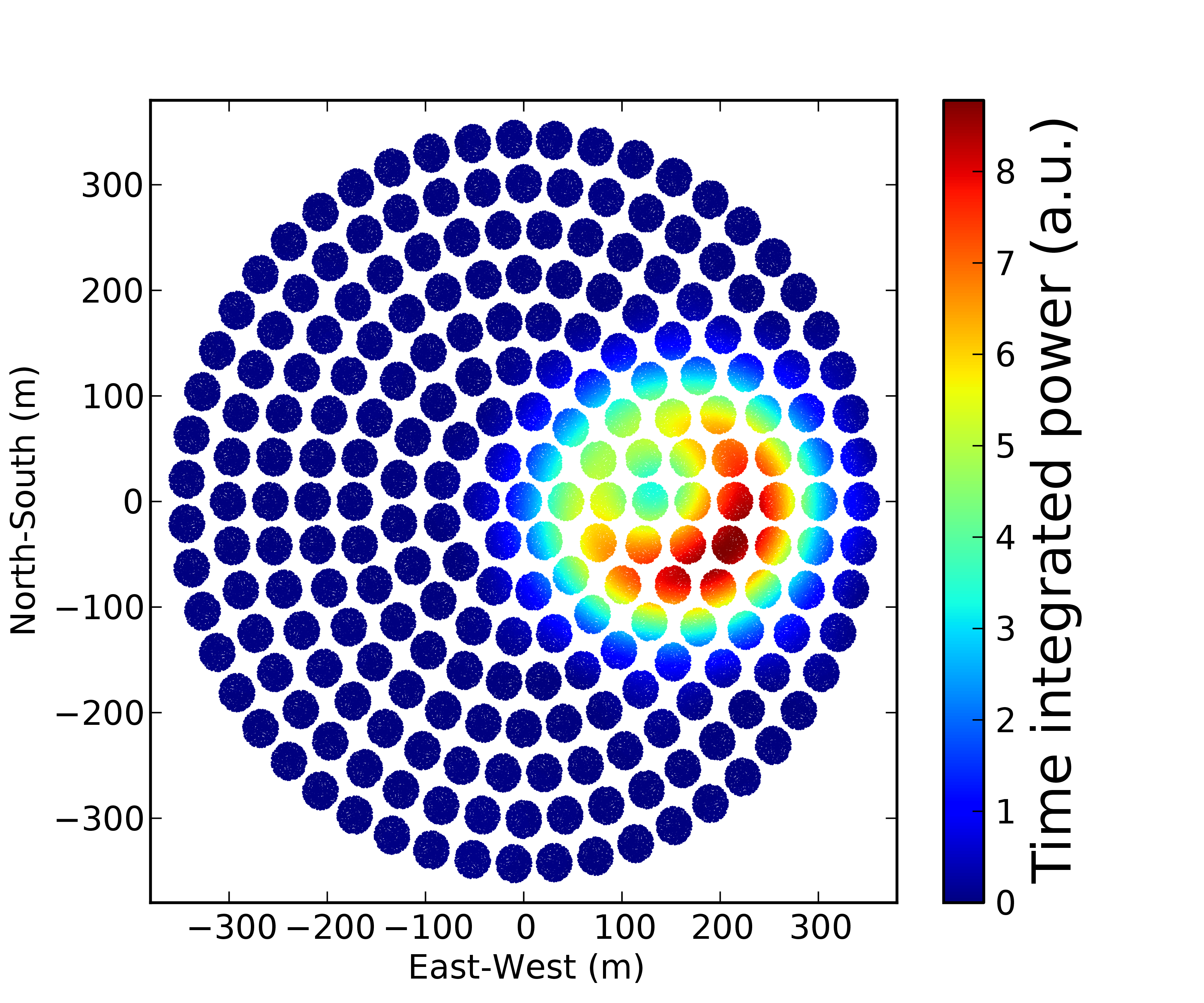}
\caption{CoREAS simulation of the radio emission footprint sampled 
with LOFAR (top-left) and SKA-low (bottom, zoomed-in at top-right). Each 
point corresponds to a measurement with an individual dual-polarized antenna. The zenith angle of the air 
shower is 30$^{\circ}$ and the energy corresponds to $10^{18}$~eV. 
Even for an ideal core position of the air shower, the LOFAR measurement provides only a very incomplete sampling of the radio 
signal. In comparison, the SKA-low sampling is extremely homogeneous 
and detailed, irrespective of the core position within the antenna array. Also note 
the appearance of a Cherenkov ring in the SKA-low measurement, which is 
due to the measurement of higher-frequency components up to 
350~MHz.\label{fig:simcomparison}}
\end{figure}

We have performed initial simulation studies based on CoREAS 
\cite{coreas} simulations to assess the improvement in the 
measurement of individual air showers with SKA-low as compared to 
LOFAR. In Fig.\ \ref{fig:simcomparison} we illustrate the amount of 
detail that a measurement with SKA-low (bottom, zoomed-in at top-right) will record as compared 
with LOFAR (top-left). The homogeneous and dense sampling of the radio 
emission in SKA-low will enable reconstruction of the depth of shower 
maximum with an estimated average uncertainty of less than 
10~g/cm$^{2}$. The widely-used fluorescence detection technique 
currently provides depth of shower maximum measurements with a  
resolution of $\sim 20$~g/cm$^{2}$, with a duty-cycle of 
roughly 10\%. SKA-low can thus be expected to provide significantly improved 
resolution of the depth of shower maximum, including systematic 
uncertainties, and with near-100\% duty-cycle. The 
high-frequency data available in SKA-low measurements will in addition provide 
further information such as the diameter of the Cherenkov ring and 
a measurement of inhomogeneities in the shower cascade.

\section{Conclusion}

With moderate engineering changes, the Square Kilometre Array can be enabled for the detection of 
extensive air showers. Its approximately 60,000 antennas in a fiducial 
area of roughly 1~km$^{2}$ will allow studies of individual  
air showers at an unprecedented level of detail. We expect an average 
uncertainty on the determination of the depth of shower maximum of 
less than 10~g/cm$^{2}$, significantly lower than the resolution of 
the widely-used fluorescence technique, and additionally information 
on fine details of the air shower longitudinal profile. Along with a precise determination of the 
cosmic ray energy, SKA-low will thus be a powerful instrument to study the 
mass composition in the energy range from $\gtrsim 10^{16}$ to 
$\gtrsim 10^{18}$~eV. Among its science goals will be the study of the 
transition from Galactic to extragalactic cosmic rays, particle 
interactions at energies beyond the reach of the LHC, and studies of 
thunderstorm physics, including potential links to cosmic rays.


\begin{thebibliography}{99}

\bibitem{huegereview} T.~Huege: \emph{Braz.\ J.\ Phys.} \textbf{44} (2014) 520
\bibitem{schulzaera} J.~Schulz for the Pierre Auger Coll.: \emph{this issue} (2015), id \#615
\bibitem{lofar} P.~Schellart, et al.: \emph{Astron.\ Astroph.} \textbf{560} (2013) A98
\bibitem{coreas} T.~Huege, M.~Ludwig \& C.W.~James, \textbf{AIP Conf.\ Proc.} \textbf{1535} (2012) 128
\bibitem{lofarxmax} S.~Buitink, et al., \emph{Phys.\ Rev.\ D.} \textbf{90} (2014) 082003
\bibitem{ska-lunar} C.W.~James, et al: \emph{this issue} (2015) id \#291 
\bibitem{ulrich} R.~Ulrich, et al., \textbf{Phys.\ Rev.\ D} \textbf{83} (2011) 054026
\bibitem{AugerCC} P.~Abreu, et al.\ (Pierre Auger Coll.), \textbf{Phys.\ Rev.\ Lett.} \textbf{109} (2012) 062002
\bibitem{lopesthunder} S.~Buitink, et al.\ (LOPES Coll.), \emph{A\&A} \textbf{467} (2007) 385
\bibitem{lofarthunder} P.~Schellart, et al: \emph{Phys.\ Rev.\ Lett.} \textbf{114} (2015) 165001
\bibitem{RunawayBreakdown} A.V.~Gurevich \& A.N.~Karashtin, \emph{Phys.\ Rev.\ Lett.} \textbf{110} (2013) 185005
\bibitem{LightningInception} A.~Dubinova, et al., \emph{Phys.\ Rev.\ Lett.} \textbf{115} (2015) 015002
\bibitem{zhaires} J.~Alvarez-Mu\~niz, W.R.~Carvalho \& E.~Zas, \emph{Astropart.\ Phys.} \textbf{35} (2012) 325
\bibitem{kascade} T.~Antoni et al. (KASCADE Coll.): \emph{Nucl.\ Instr.\ Meth.\ A} \textbf{513} (2003) 490

















 





 
 




\end{thebibliography}
\end{document}